\documentclass[a4paper, 10pt, onecolumn]{article}      

\usepackage{graphics} 
\usepackage{epsfig} 
\usepackage{amsmath} 
\usepackage{amssymb}  
\usepackage{theorem}
\newtheorem{theorem}{Theorem}
\newtheorem{remark}{Remark}
\newtheorem{proof}{Proof}

\title{\LARGE \bf
Robust control tools for traffic monitoring in TCP/AQM networks
}

\author{Yassine Ariba$^{\ast\dag}$, Fr\'ed\'eric Gouaisbaut$^{\ast\dag}$, Sandy Rahme$^{\ast\dag}$ and Yann Labit$^{\ast}$ 
\thanks{ Universit\'e de Toulouse; UPS, 118 Route de Narbonne, F-31062 Toulouse, France.}%
\thanks{ LAAS; CNRS;  7, avenue du Colonel Roche, F-31077 Toulouse, France.
        {\tt\small yariba@laas.fr}}%
}

\date{January 2009}

\begin{document}

\maketitle
\thispagestyle{empty}
\pagestyle{empty}

\begin{abstract}
Several studies have considered control theory tools for traffic control in communication networks, as for example the congestion control issue in IP (Internet Protocol) routers. In this paper, we propose to design a linear observer for time-delay systems to address the traffic monitoring issue in TCP/AQM (Transmission Control Protocol/Active Queue Management) networks. Due to several propagation delays and the queueing delay, the set TCP/AQM is modeled as a multiple delayed system of a particular form. Hence, appropriate robust control tools as quadratic separation are adopted to construct a delay dependent observer for TCP flows estimation. Note that, the developed mechanism enables also the anomaly detection issue for a class of DoS (Denial of Service) attacks. At last, simulations via the network simulator NS-2 and an emulation experiment validate the proposed methodology.

\end{abstract}

\section{Motivations and Contributions}
~\indent Internet is becoming the major communication network. It allows an increasing number of activities, ranging from web browsing, file exchanges to on-line games or IP telephony. Because of its increasing popularity, traffic monitoring tools have to be embedded into the network to supervise communications to ensure QoS (Quality of Service) or even to avoid security breaches. Two techniques can be used:\\
~\indent Active monitoring \cite{Pra03} consists of generating probes into the network, and then to observe the impact of network components and protocols on traffic: loss rate, delays, RTT (Round Trip Time), capacity... However, since an additional traffic (probes) is injected into the network, the major drawback is the disturbance induced by such traffic (it inevitably affects the current traffic). Intrusiveness of probe traffic is thus one of the key features which active monitoring tools have to care about.\\
~\indent Secondly, passive monitoring  \cite{Cle00} refers to network measurements with appropriate devices located at some relevant point in the network. Passive monitoring is performed on the capture of traffic and off-line estimate networks features. It provides a non intrusive method but not enough reactive.   
\\~\indent Regarding the security problems, network anomalies typically refer to circumstances when network operations deviate from the expected behavior. Network anomalies can be roughly classified into two categories. The first category is related to network failures and performance problems (like file server failures, broadcast storms, etc...). The second major category of network anomalies is security-related problems (like DoS or DDoS detections) in detecting active security threats. A variety of tools for anomaly detection are mainly based on data packet signatures (i.e. specific formats of packages, packet headers) and the use of statistical profiles of the traffic. The natural variability of the traffic \cite{Park96} produces important fluctuations of these measurements, inducing thus several false positives (false alarms) and false negatives (missed detections). Some studies have taken into account a richer form of the statistical structure of the traffic (correlation, spectral density ...) to design IDS or ADS (Intrusion or Anomaly Detection System) \cite{Hus03}, \cite{lak04}.\\
~\indent In this paper, we propose to address the traffic monitoring issue in networks with the design of an observer. First, a dynamical model which describes the TCP flow rates behavior as well as a class of anomalies is introduced. Then, robust control tools, especially quadratic separation, are used to derive a convergence condition for the time delay observer. Basically, the observer, embedded at a router, uses the queue length measurement of the buffer to reconstruct the whole state composed of flow rates. However, this latter being related to the linearized model of TCP, traffic has to be regulated around an equilibrium point to ensure the validity of the observer model and a congestion control mechanism (as AQM, Active Queue Management) is thus required. Next, the model is extended in order to detect a class of anomalies from the second category (attacks). Note that the proposed methodology allows on-line and non-intrusive monitoring (as active monitoring but without injecting probes into the network). Even if our study focuses on specific and static networks as explained in the next section, it shows encouraging results. 
\\~\indent The paper is organized as follows. The problem statement introducing the model of a network supporting TCP and the AQM congestion control is presented in the second section. Then, the third part is dedicated to the design of an observer for the estimation of data flow rates as well as anomaly detection. The fourth section shows an illustrative example of the proposed theory using NS-2 simulations and emulations. Finally, the fifth section concludes the paper and proposes future works.\\

\section{NETWORK DYNAMICS}

\subsection{Fluid-flow model of TCP}
\begin{figure}
       \centerline{\includegraphics[width=7.5cm,height=5.5cm]{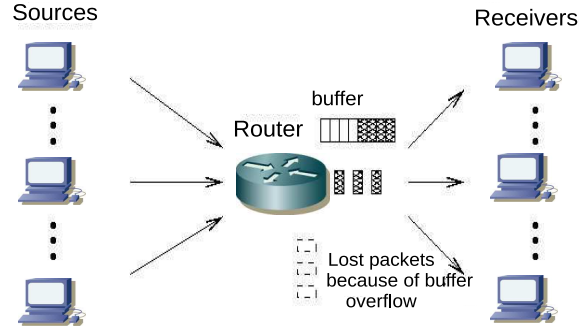}}
       \caption{Network topology}
       \label{topologie}
       \end{figure}

~\indent  This section is devoted to the introduction of the network model that describes the traffic behavior. In this paper, we consider networks consisting of a single router and $N$ heterogeneous TCP sources. By heterogeneous, we mean that each source is linked to the router with different propagation times (see Figure \ref{topologie}).\\
~\indent Since the bottleneck is shared by $N$ flows, TCP applies the congestion avoidance algorithm to avoid the network saturation \cite{Jac88}. Following the AIMD ({\it Additive-Increase Multiplicative-Decrease}) mechanism, the congestion window of TCP sources varies according to the network load state (packet losses and delays). Hence, various deterministic fluid-flow models have been developed (see \cite{Low02}, \cite{Mis00} and \cite{Sri04} and references therein) to describe the behavior of the transmission protocol.
\begin{figure}
        \centerline{\includegraphics[=6cm,height=6cm]{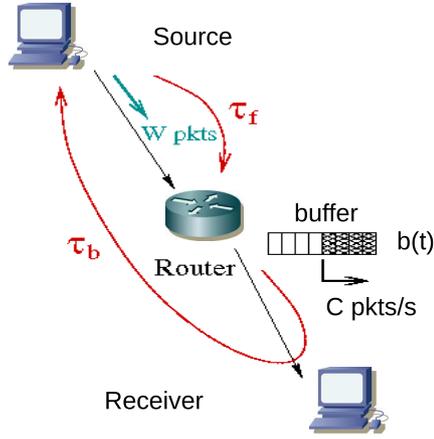}}
       \caption{A single connection}
       \label{topologie4}
\end{figure}
While many studies dealing with network control in the automatic control theory framework consider the model proposed by \cite{Mis00}, we use a more accurate one, introduced in \cite{Low02} and described by  (\ref{modelNL}) which takes into account the forward and backward delays. The model and notations are as follow:
\begin{equation}
  \label{modelNL}
\left\{\begin{array}{rl}
\dot{W}_i(t)&=\frac{W_i(t-\tau_i)}{\tau_i(t-\tau_i)}(1-p_i(t-\tau_i^b))\frac{1}{W_i(t)}\\&~~~~-\frac{W_i(t-\tau_i)}{\tau_i(t-\tau_i)}\frac{W_i}{2}p_i(t-\tau_i^b),\\
\dot{b}(t)&=-c+\sum_{i=1}^N\eta_i\frac{W_i(t-\tau_i^f)}{\tau_i(t-\tau_i^f)},\\
\tau_i&=\frac{b(t)}{c}+T_{p_i}=\tau_i^f+\tau_i^b,
\end{array} \right.
\end{equation}
where  $W_i(t)$ is the congestion window size of the source $i$, $b(t)$ is the queue length of the buffer at the router, $\tau_i$ is the RTT perceived by the source $i$. This latter quantity can be decomposed as the sum of the forward and backward delays ($\tau_i^f$ and $\tau_i^b$), standing for, respectively, the trip time from the source $i$ to the router (the one way) and from the router to the source via the receiver (the return) (see Figure \ref{topologie4}). $c$, $T_{p_i}$ and $N$ are parameters related to the network configuration and represent, respectively, the link capacity, the propagation time of the path taken by the connection $i$ and the number of TCP sources. $\eta_i$ is the number of sessions established by source $i$. The signal $p_i(t)$ corresponds to the dropping probability of a packet at the router buffer. Note that the network variables mentionned above in model (\ref{modelNL}) are considered as mean values \cite{Low02} (for instance, $W_i(t)$ represents actually the average congestion window size).\\~\indent

In this paper, the objective is to develop a method which computes, at the router and during congestion, an estimation of the different flow rates passing through it. The congestion window $W_i$ does not provide a relevant index of the traffic intensity since it only refers to the amount of data sent by the source at a given instant. Consequently, additional frequent measures of the corresponding $RTT$ are required. Hence, we propose to reformulate the model (\ref{modelNL}) such that the state vector is expressed in terms of aggregate flows instead of congestion windows. To this end, rates of each flow $x_i$, expressed as $x_i(t)=\frac{W_i(t)}{\tau_i(t)}$, will be considered. The dynamic of this new quantity becomes of the form $\dot{x_i}(t)=\frac{d}{dt}\left(\frac{W_i(t)}{\tau_i(t)}\right)=\frac{\dot{W_i}(t)-x_i(t)\dot{\tau_i}(t)}{\tau_i(t)}$. Based on the expressions of $\dot{W}(t)$, $\dot{b}(t)$, $\tau_i(t)$ (see equation (\ref{modelNL})) and $\dot{\tau}(t)=\frac{\dot{b}(t)}{c}$, a new model of the TCP behavior is derived

\begin{equation}
\label{modelNL2}
\left\{\begin{array}{rl}
\dot{x}_i(t)&=\frac{x_i(t-\tau)}{x_i(t)\tau(t)^2}(1-p(t-\tau ^b))-\frac{x_i(t-\tau)x_i(t)}{2}p(t-\tau^b)\\&~~~~+\frac{x_i(t)}{\tau(t)}-\frac{x_i(t)}{\tau(t)c}\sum_i\eta_ix_i(t-\tau^f_i)\\
\dot{b}(t)&=-c+\sum_{i=1}^N\eta_ix_i(t-\tau_i^f)
\end{array} \right.\!\!\!.
\end{equation}

\subsection{AQM for congestion control}

\begin{figure}
       \centerline{\includegraphics[height=.25 \textheight, width=8cm]{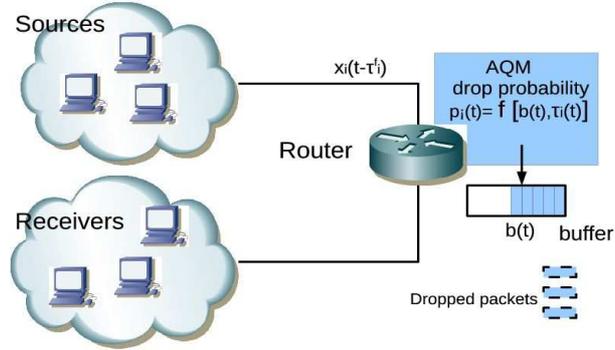}}
       \caption{Implementation of an AQM}
       \label{schema_block_struct}
\end{figure}

To achieve high efficiency and high reliability of communications in computer networks, many investigations have been done regarding the congestion control issue. Since the congestion window size of the transmission protocol depends on packet losses (specified by $p_i(t)$), a proposal was to use this feature in order to control the source sending rates. Hence, a mechanism, called AQM ({\it Active Queue Management}, see Figure \ref{schema_block_struct}), has been developed to provoke losses avoiding then severe congestion, buffer overflow, timeout... This strategy allows the regulation of TCP flows with an implicit control (or explicit if the ECN, {\it Explicit Congestion Notification}, protocol is enabled). Various AQM have been proposed in the literature such as Random Early Detection (RED) \cite{Flo93}, Random Early Marking (REM) \cite{Ath00}, Adaptive Virtual Queue (AVQ) \cite{Sri04} and many others \cite{Ryu04}. Their performances have been evaluated in \cite{Ryu04} and empirical studies have shown their effectiveness. Recently, significant studies initiated by \cite{Hol02} have redesigned AQMs using control theory and $P$, $PI$ have been developed in order to cope with the packet dropping problem. Then, using dynamical model developed by \cite{Mis00}, many researches have been devoted to deal with congestion problem in a control theory framework (for examples see \cite{Lab07b}, \cite{Kim06}, \cite{Tar05} and references therein).

~\indent So, AQM supports TCP for congestion control and regulates the queue length of the buffer as well as flow rates around an equilibrium point \cite{Lab07b}, \cite{Kim06}, \cite{Hol02}. An efficient control allows thus to approximate the TCP dynamics (\ref{modelNL2}) as a linear model (\ref{forme_can}) around an equilibrium point (\ref{point_eq}). Our work focuses on traffic monitoring at a router with a static topology ($N$ and $\eta_i$ are constant). Moreover, for the mathematical tractability, we make the usual assumption \cite{Low02}, \cite{Hol02}, \cite{Kim06} that all delays ($\tau_i$, $\tau_i^f$ and $\tau_i^b$) are time invariant when they appear as arguments of variables (for example $x_i(t-\tau_i(t))\equiv x_i(t-\tau_i)$). This latter assumption is valid as long as the queue length remains close to its equilibrium value and when the queueing delay is smaller than propagation delays. Defining an equilibrium point
\begin{equation}
  \label{point_eq}
\left\{\begin{array}{rl}
\tau_{i_0}&=T_p+b_0/c\\
\dot{b}(t)&=0~\Rightarrow~\sum_{i=1}^N\eta_ix_{i_0}=c\\
\dot{x}_i(t)&=0~\Rightarrow~p_{i_0}=\frac{2}{2+(x_{i_0}\tau_{i_0})^2}
\end{array} \right.,
\end{equation}
model (\ref{modelNL2}) can be linearized to obtain:
{\small
\begin{equation}
\label{forme_can}
\begin{aligned}
\left[\begin{array}{c}
\dot{x}_1(t)\\\vdots\\\dot{x}_N(t)\\\dot{b}(t)\end{array}\right]=&
A\left[\begin{array}{c}
\delta x_1(t)\\\vdots\\ \delta x_N(t)\\\delta b(t)\end{array}\right]+
A_d
\left[\begin{array}{c}
\delta x_1(t-\tau_1^f)\\\vdots\\\delta x_N(t-\tau_N^f)\\\delta b(t)\end{array}\right]\\&~~+
B\left[\begin{array}{c}
\delta p_1(t-\tau_1^b)\\ \vdots\\ \delta p_N(t-\tau_N^b)\end{array}\right]
\end{aligned}
\end{equation}}
where $\delta x_i \doteq x_i-x_{i_0}$, $\delta b \doteq b-b_0$ and $\delta p_i
\doteq p_i-p_{i_0}$ are the state variations around the equilibrium point (\ref{point_eq}). Matrices of the equation (\ref{forme_can}) are defined by
\begin{equation*}
\begin{aligned}
A=&\left[\begin{array}{cccc}a_1&\sf{0}&\sf{0}&h_1\\\sf{0}&\ddots&\sf{0}&\vdots\\\sf{0}&\sf{0}&a_N&h_N\\\sf{0}&\sf{0}&\sf{0}&\sf{0}\end{array}\right],~~B=\left[\begin{array}{ccc}e_1&0&0\\0&\ddots&0\\0&0&e_N\\0&0&0\end{array}\right]\\
A_d=&\left[\begin{array}{cccc}f_1\eta_1&\ldots&f_1\eta_N&\sf{0}\\\vdots&\vdots&\vdots&\sf{0}\\f_N\eta_1&\ldots&f_N\eta_N&\sf{0}\\\eta_1&\ldots&\eta_N&\sf{0}\end{array}\right],
\end{aligned}
\end{equation*}
with $a_i=-\frac{1-p_{i_0}}{x_{i_0}\tau^2_{i_0}}-\frac{x_{i_0}p_{i_0}}{2}$, $h_i=-\frac{2(1-p_{i_0})}{c\tau^3_{i_0}}$, $f_i=-\frac{x_{i_0}}{\tau_{i_0}c}$ and $e_i=-\frac{1}{\tau_{i_0}^2}-\frac{x_{i_0}^2}{2}$.
Remark that a multiple time delays system \eqref{forme_can} is obtained with a particular form since each component of the state vector is delayed by a different quantity related to the communication path.

\section{OBSERVER FOR TRAFFIC MONITORING}

\subsection{Preliminaries}

\begin{figure}
       \centerline{\includegraphics[height=.08 \textheight]{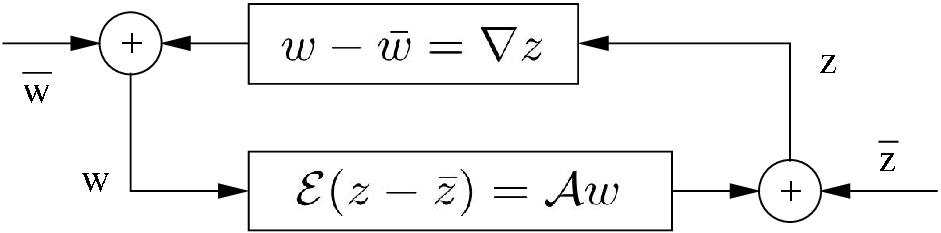}}
       \caption{An interconnected system}
       \label{feedbacksystem}
\end{figure}

First, and before designing the oberver, it is necessary to introduce the following theorem \cite{Pea07} that provides stability condition for interconnected systems as illustrated in Figure \ref{feedbacksystem}. This result is then used to cope with the delayed part of (\ref{forme_can}) and to provide conditions for the convergence of the observer state to (\ref{forme_can}).
\begin{theorem}
\label{theorem}
Given two possibly non-squared matrices $\mathcal{E}$, $\mathcal{A}$ and an uncertain matrix
 $\nabla$ belonging to a set $\Xi$. The uncertain system represented by Figure \ref{feedbacksystem} is stable for all matrices $\nabla\in\Xi$ if and only if there exists a matrix
 $\Theta=\Theta^*$ satisfying conditions
\begin{align}
\label{cond-theorem1}
\left[\begin{array}{cc}\mathcal{E}&-\mathcal{A}\end{array}\right]^{\perp*}\Theta
\left[\begin{array}{cc}\mathcal{E}&-\mathcal{A}\end{array}\right]^{\perp}>\sf{0}\\
\label{cond-theorem2}
\left[\begin{array}{cc}\sf{1}&\nabla^*\end{array}\right]\Theta
\left[\begin{array}{c}\sf{1}\\ \nabla\end{array}\right]\leq\sf{0}.
\end{align}
\end{theorem}
The considered feedback system having the same form of Figure \ref{feedbacksystem} is a linear equation connected to a linear uncertainty $\nabla$. This result comes from robust control theory using the quadratic separation tools \cite{Iwa98}. The second inequality (\ref{cond-theorem2}) is constructed based on some knowledge about the uncertain matrix $\nabla$ (for instance upperbounds, convex hull). Then, the first one (\ref{cond-theorem1}) is solved to assess the stability of the interconnection. Previous works \cite{Gou06a} have shown that such a framework provides convenient tools and a good insight into time delay systems stability issue. In that case, the delay system is represented as in Figure \ref{feedbacksystem} where $\nabla$ consists of some appropriate operators related to the delay.\\

~\indent In the next part, Theorem \ref{theorem} leads to conceive an observer that tracks the state of the multiple time delays system (\ref{forme_can}).

\subsection{Design of the observer}

\begin{figure}
       \centerline{\includegraphics[height=6.5cm, width=8.5cm]{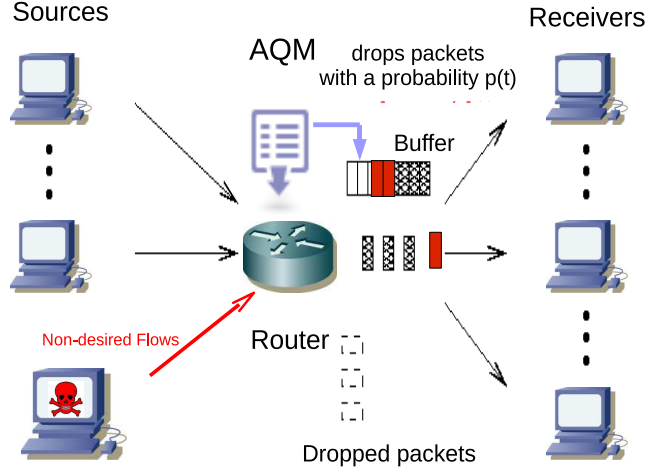}}
       \caption{Introduction of an additional non-TCP traffic as anomaly}
       \label{topologie3}
\end{figure}

Consider a network as illustrated in Figure \ref{topologie} consisting of $N$ TCP pairs, the traffic dynamic regulated by an AQM can be modeled around the equilibrium point as (see (\ref{forme_can}))
\begin{equation}
\label{forme_can2}
\left\{\begin{array}{rcl}
\dot{x}(t)&=&Ax(t)+A_dx_d(t)+Bu(t)\\
y(t)&=&Cx(t)
\end{array}\right.
\end{equation}
where
\begin{equation*}
\begin{array}{c}
x(t)=\left[\begin{array}{c}\delta x_1(t)\\ \vdots \\ \delta x_N(t) \\ \delta b(t)\end{array}\right],~
x_d(t)=\left[\begin{array}{c}\delta x_1(t-\tau_1^f)\\ \vdots \\ \delta x_N(t-\tau_N^f)\\ \delta b(t)\end{array}\right],\\[0.3cm]
u(t)=\left[\begin{array}{c}\delta b(t-\tau_1^b)\\ \vdots \\ \delta b(t-\tau_N^b)\end{array}\right],~C=\left[\begin{array}{cccc}0&\ldots&0&1\end{array}\right].
\end{array}
\end{equation*}
and $y(t)$ is the measured output $i.e.$ the queue length at the router. In order to take into account extra traffic or non-modeled traffic (for example, traffics coming from applications over UDP protocol, see Figure \ref{topologie3}), an additional signal $d(t)$ should be added to the queue dynamic (second equation in (\ref{modelNL2})):
\begin{equation*}
\dot{b}(t)=-c+d(t)+\sum_{i}\eta_ix_i(t-\tau_i^f).
\end{equation*}
This signal represents flows that pass through the router and fill up the buffer $b(t)$ in addition to the expected traffic ($N$ TCP connections). Notice that this feature can be used to model anomalies or DoS attacks ({\it Denial of Service}, \cite{cert}).
 In this paper, we consider some class of anomalies that are CBR ({\it Constant Bit Rate}) based applications which can be modeled as piecewise-constant functions. Such applications are met in streaming applications, video conferencing, telephony (voice services). Furthermore, the same modeling can also be used for some class of attacks \cite{Luo05} as traditional {\it flooding-based DoS} (for example $Shrew$) or {\it PDoS} (see \cite{Luo05} and references therein). Consequently, assuming that $d(t)$ is a piecewise-constant function, we propose to consider now the following augmented system which embeds the anomaly feature:
\begin{equation}
\label{augmented-model}
\left\{\begin{array}{rcl}
\dot{\tilde{x}}(t)&=&\bar{A}\tilde{x}(t)+\bar{A}_d\tilde{x}_d(t)+\bar{B}u(t)\\
\tilde{y}(t)&=&\bar{C}\tilde{x}(t)
\end{array}\right.
\end{equation}
where $\tilde{x}(t)=\left[\begin{array}{c}x(t)\\ \hline d(t) \end{array}\right],~\bar{C}=\left[\begin{array}{c|c}C&0 \end{array}\right]$,
\begin{equation*}
\begin{array}{c}
\!\!\!\bar{A}=\left[\begin{array}{c|c}A&\begin{array}{c}0\\ \vdots \\ 0 \\1\end{array}\\ \hline \begin{array}{ccc}0& \ldots &0\end{array}&0 \end{array}\right]\!,~\tilde{x}_d(t)=\left[\begin{array}{c}x_d(t)\\ \hline d(t) \end{array}\right],\\
\!\!\!\!\bar{B}=\left[\begin{array}{c}B\\ \hline \sf{0}_{1\times N} \end{array}\right],~\bar{A}_d=\left[\begin{array}{c|c}\!\!A_d\!\!&\begin{array}{c}0\\ \vdots \\ 0 \end{array}\\ \hline \!\!\begin{array}{ccc}0& \ldots & 0 \end{array}\!\!&0 \end{array}\right]
.\\
\end{array}
\end{equation*}
Let construct an observer for the augmented system (\ref{augmented-model}) defined by:
\begin{equation}
\label{observer}
\dot{\hat{x}}(t)=\bar{A}\hat{x}(t)+\bar{A}_d\hat{x}_d(t)+\bar{B}u(t)+L\left(y-\bar{C}\hat{x}(t)\right)
\end{equation}
where $\hat{x}(t)$ is the observer state and $L$ is the observer gain. This latter matrix has to be designed such that $\hat{x}(t)$ converges to $\tilde{x}(t)$. Notice that the pair $(\bar A+\bar A_d,\bar C)$ is observable which implies that there exists an observer (depending eventually on the delay) allowing the reconstruction of the states of system (\ref{augmented-model}).
\begin{theorem}
\label{theorem2}
If there exists $(N+2)\times (N+2)$ positive definite matrices $P$, $Q_{i}$ and $S_{i}$ for $i=\{1,...,N\}$ and a matrix $X\in\sf{R}^{(N+2)\times 1}$ such that the following inequality holds
\begin{equation}
\label{cond-theorem}
\left[\begin{array}{cccc}
\Xi_1+\Xi_3&Y&\ldots&Y\\
Y^T&\frac{1}{\tau_1^{f^2}}S_1&&\sf{0}\\
\vdots&&\ddots&\\
Y^T&\sf{0}&&\frac{1}{\tau_N^{f^2}}S_N
\end{array}\right]>\sf{0}
\end{equation}
with
\begin{align}
\label{def-xi1}
\Xi_1=&\left[\begin{array}{cccc}\Psi&-P\bar{A}_{d_1}&\ldots&-P\bar{A}_{d_N}\\-\bar{A}^T_{d_1}P&Q_1&&\sf{0}\\ \vdots &&\ddots&\\-\bar{A}^T_{d_N}P&\sf{0}&&Q_N \end{array}\right],\\
\label{def-xi3}
\Xi_3=&\sum_{i=1}^NM_i(2P-S_i)M_i^T,\\
Y=&\left[\begin{array}{cccc}(P\bar A-X\bar C)& P\bar A_{d_1}& \ldots& P\bar A_{d_N}\end{array}\right]^T\\
\Psi=&-P\bar{A}-\bar{A}^TP+X\bar C+\bar C^T X^T-\sum_{i=1}^NQ_i,\\
\label{def-Mi}
M_i=&\left[\begin{array}{c}-\sf{1}_{N+2}\\ \sf{0}_{(N+2)(i-1)\times(N+2)} \\ \sf{1}_{N+2} \\ \sf{0}_{(N-i)(N+2)\times(N+2)}\end{array}\right],
\end{align}
then system (\ref{observer}) is an observer for system (\ref{augmented-model}), $i.e$ $\hat{x}(t)$ converges asymptotically to $\tilde{x}(t)$. The observer gain $L$ is given by $L=P^{-1}X$.
\end{theorem}
\begin{proof}
In order to prove the asymptotic convergence of $\hat{x}(t)$ to $\tilde{x}(t)$, let us define the error between the state of (\ref{augmented-model}) and the one of the observer (\ref{observer}): $e(t)=\tilde{x}(t)-\hat{x}(t)$. We aim to make sure that the error $e(t)$ converges toward zero. Hence, the first problem can be recast  as the stability issue of system
\begin{equation}
\label{error}
\dot{e}(t)=\left(\bar{A}-L\bar{C}\right)e(t)+\bar{A}_de_d(t).
\end{equation}
where $e_d(t)=\tilde{x}_d(t)-\hat{x}_d(t)$. System (\ref{error}) is then rewritten as 
\begin{equation}
\label{error2}
\dot{e}(t)=\mathbb{A}e(t)+\sum_{i=1}^{N}\bar{A}_{d_i}e(t-\tau_i^f)
\end{equation}
with $\mathbb{A}=\left(\bar{A}-L\bar{C}\right)$,{\small
\begin{equation*}
\bar{A}_{d_i}=\eta_i\left[\begin{array}{c|c|c}\sf{0}_{(N+2)\times(i-1)}&\begin{array}{c}f_1\\ \vdots\\ f_N\\1\\0\end{array}&\sf{0}_{(N+2)\times(N-i+2)}\end{array}\right].
\end{equation*}}
Next, transforming the system (\ref{error2}) as an interconnected system of the form of Figure \ref{feedbacksystem}, Theorem \ref{theorem} may be applied to derive the stability condition. System (\ref{error2}) is thus expressed as the interconnection of
\begin{equation}
\label{nabla}
w(t)=
\underbrace{\left[\begin{array}{ccc}s^{-1}\sf{1}_{N+2}&\sf{0}&\sf{0}\!\\ \sf{0}&\mathcal{D}\otimes\sf{1}_{N+2}&\sf{0}\!\\ \sf{0} &\sf{0}&(\sf{1}-\mathcal{D})s^{-1}\otimes\sf{1}_{N+2}\!\end{array}\right]}_{\nabla}z(t)\!\!
\end{equation}
and equation (\ref{EA})
\begin{table*}
\begin{equation}
\label{EA}
\underbrace{\left[\begin{array}{c}
\begin{array}{cc}\sf{1}_{(N+2)(N+1)}&\sf{0}_{(N+2)(N+1)\times (N+2)N}\end{array}\\ \hline
\begin{array}{c|cc}\begin{array}{c}-\sf{1}_{(N+2)}\\ \vdots \\ -\sf{1}_{(N+2)}\end{array}&\sf{0}_{(N+2)N}&\sf{1}_{(N+2)N}\end{array}\\ \hline
\sf{0}_{(N+2)N\times(N+2)(2N+1)}
\end{array}\right]}_{\mathcal{E}}\underbrace{\left[\begin{array}{c}\dot{e}(t)\\ e(t)\\ \vdots\\ e(t)\\ \dot{e}(t) \\ \vdots\\ \dot{e}(t)\end{array}\right]}_{z(t)}=
\underbrace{\left[\begin{array}{c}
\begin{array}{ccccc}\mathbb{A}&\bar{A}_{d_1}&\ldots&\bar{A}_{d_N}&\sf{0}\end{array}\\ \hline
\begin{array}{c|c}\begin{array}{c}\sf{1}_{(N+2)}\\ \vdots \\ \sf{1}_{(N+2)}\end{array}&\sf{0}_{N(N+2)\times 2N(N+2)}\end{array}\\ \hline
\sf{0}_{(N+2)N\times(N+2)(2N+1)}\\ \hline
\begin{array}{c|cc}\begin{array}{c}\sf{1}_{(N+2)}\\ \vdots \\ \sf{1}_{(N+2)}\end{array}&-\sf{1}_{(N+2)N}&-\sf{1}_{(N+2)N}\end{array}\\ 
\end{array}\right]}_{\mathcal{A}}\underbrace{\left[\begin{array}{c}e(t)\\ e(t-\tau_1^f) \\ \vdots\\ e(t-\tau_N^f)\\ e(t)-e(t-\tau_1^f)\\ \vdots \\ e(t)-e(t-\tau_N^f)\end{array}\right]}_{w(t)}
\end{equation}
\end{table*}
where $\mathcal{D}=diag\left(e^{-\tau_1^fs},...,e^{-\tau_N^fs}\right).$\\
~\indent First, it can be proved that the separator (\ref{separateur}) satisfies the inequality (\ref{cond-theorem2}) according to the matrix $\nabla$ defined as (\ref{nabla}) (proof is omitted because of the space limitation, it is an extension of \cite{Gou06a} to the case of multiple delays).
\begin{equation}
\label{separateur}
\Theta=\left[\begin{array}{c|c}\Theta_{11}&\Theta_{12}\\ \hline \ast &\Theta_{22}\end{array}\right]
\end{equation}
with
\begin{equation*}
\begin{aligned}
\Theta_{11}=& ~diag\left(\sf{0}_{N+2},-Q_1,\ldots,-Q_N,-R_1\tau_1^{f^2},\ldots,-R_N\tau_N^{f^2}\right)\\
\Theta_{12}=& ~diag\left(-P,\sf{0}_{2N(N+2)}\right)\\
\Theta_{22}=& ~diag\left(\sf{0}_{N+2},Q_1,\ldots,Q_N,R_1,\ldots,R_N\right)
\end{aligned}
\end{equation*}
$P$, $Q_i$ and $R_i$ $\forall i\in[1,N]$ are positive definite matrices. So, system (\ref{error}) is stable if the inequality (\ref{cond-theorem1}) with $\mathcal{E}$ and $\mathcal{A}$ defined as (\ref{EA}) is verified. Some algebraic calculations show this latter is of the form
\begin{equation}
\label{proof-cond}
\bar \Xi_1-\Xi_2+\bar\Xi_3>\sf{0}
\end{equation}
with
\begin{equation*}
\begin{aligned}
\bar\Xi_1=&\left[\begin{array}{cccc}-P\mathbb{A}-\mathbb{A}^TP-\sum_iQ_i&-P\bar{A}_{d_1}&\ldots&-P\bar{A}_{d_N}\\-\bar{A}^T_{d_1}P&Q_1&&\sf{0}\\ \vdots &&\ddots&\\-\bar{A}^T_{d_N}P&\sf{0}&&Q_N \end{array}\right],\\
\Xi_2=&\left[\begin{array}{c}\mathbb{A}^T\\\bar{A}^T_{d_1}\\ \vdots\\\bar{A}^T_{d_N}\end{array}\right]
\sum_{i=1}^N\tau^{f^2}_iR_i\left[\begin{array}{c}\mathbb{A}^T\\\bar{A}^T_{d_1}\\ \vdots\\\bar{A}^T_{d_N}\end{array}\right]^T,\\
\bar\Xi_3=&\sum_{i=1}^NM_iR_iM_i^T.
\end{aligned}
\end{equation*}
and $M_i$ is defined in (\ref{def-Mi}). $\Xi_2$ and $\bar\Xi_3$ are then equivalently rewritten as
\begin{equation*}
\left[\begin{array}{c}\mathbb{A}^TP\\\bar{A}^T_{d_1}P\\ \vdots\\\bar{A}^T_{d_N}P\end{array}\right]
\sum_i\tau^{f^2}_iP^{-1}R_iP^{-1}\left[\begin{array}{c}\mathbb{A}^TP\\\bar{A}^T_{d_1}P\\ \vdots\\\bar{A}^T_{d_N}P\end{array}\right]^T,
\end{equation*}
\begin{equation*}
\sum_i\left[\begin{array}{c}-P\\ \sf{0}_{(N+2)(i-1)\times(N+2)} \\ P \\ \sf{0}_{(N-i)(N+2)\times(N+2)}\end{array}\right]
P^{-1}R_iP^{-1}\left[\begin{array}{c}\ast\end{array}\right]^T
\end{equation*}
respectively. Defining $S_i=PR_i^{-1}P$ and knowing that $(P-S_i)^TS_i^{-1}(P-S_i)\geq\sf{0}$ which yields to $PS_i^{-1}P\geq 2P-S_i$, inequality $\bar\Xi_1-\Xi_2+\Xi_3> \sf{0}$ with $\Xi_3$ defined in (\ref{def-xi3}),
implies (\ref{proof-cond}). Applying a schur complement to this latter inequality and defining $X=PL$, condition (\ref{cond-theorem}) is recovered.
\end{proof}
 
\section{SIMULATION AND EMULATION}

\subsection{NS-2 simulation}

\begin{figure}[h]
       \centerline{\includegraphics[height=6cm, width=8cm]{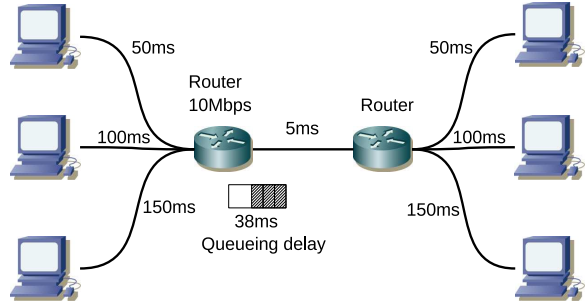}}
	\vspace*{-1cm}
       \caption{Example of a bottleneck link}
       \label{exemple}
\end{figure}

This section is dedicated to elucidate the proposed methodology through an illustrative example. As shown in Figure \ref{exemple}, a network consisting of three communicating pairs through a congested router, $i.e.$ a bottleneck, is considered. Propagation times are as illustrated and the link bandwidth is fixed to $10Mbps$, that is $2500$ packet/s considering packet size of $500$ bytes. Each of the three sources uses TCP/Reno and establishes $20$ connections generating long lived TCP flows (like FTP connections). Simulations have been performed with the network simulator NS-2 \cite{Fal02} (release 2.30) to validate the exposed theory.

 The three TCP sources share the single link and a congestion phenomenon occurs at the first router. So, to control the queue length of the buffer (avoiding then overflows), an AQM is embedded in the router. If an efficient regulation is maintained, the proposed linear observer (\ref{observer}) can be added in the router for flow monitoring. In our example, the observer have been tested over AQM gain-K \cite{Lab07b}. This latter is adjusted such that it regulates the queue length of the router to a desired level $b_0=100$ packets while the maximal buffer size is set to $400$ packets.\\
~\indent Given the topology in Figure \ref{exemple}, the previous specifications and the equilibrium point (\ref{point_eq}), the observer is then written as

\begin{figure}
       \centerline{\includegraphics[height=14cm, width=8.5cm]{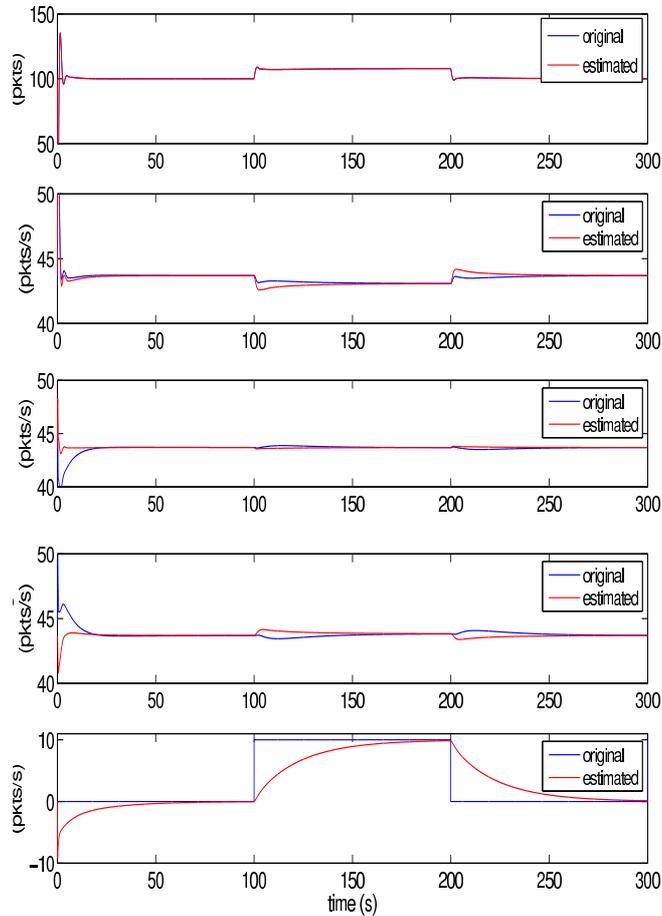}}
\vspace*{-1cm}
       \caption{Observer over gain-K: original/estimated states and anomaly detection (non linear simulation on Matlab)}
       \label{matlab}
\end{figure}

\begin{figure}
       \centerline{\includegraphics[height=14cm, width=9.5cm]{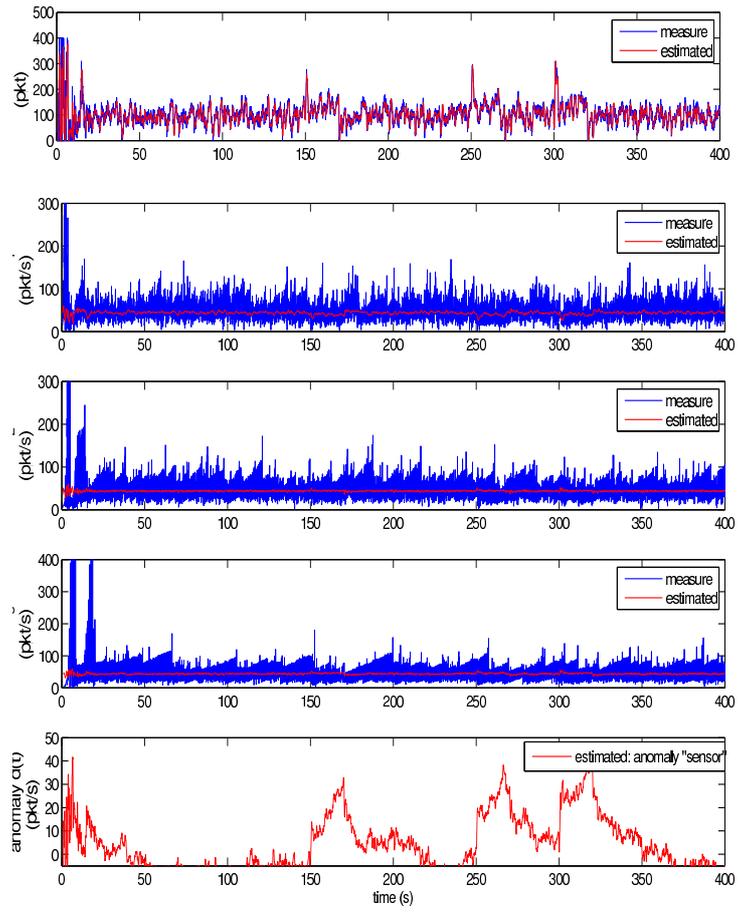}}
\vspace*{-1.2cm}
       \caption{Observer over gain-K: original/estimated states and anomaly detection (simulation on NS)}
       \label{observer_K}
\end{figure}

{\scriptsize
\begin{equation}
\label{observer-exemple}
\begin{aligned}
&\left[\begin{array}{c}\delta\dot{\hat{x}}_1(t)\\\delta\dot{\hat{x}}_2(t) \\\delta \dot{\hat{x}}_3(t)\\ \dot{\hat{b}}(t)\\ \dot{\hat{d}}(t)\end{array}\right]=\left[\begin{array}{ccccc}
-0.73\!&0\!&0\!&-0.049&0\\
0\!&-0.22\!&0\!&-0.008&0\\
0\!&0\!&-0.10\!&-0.002&0\\
0\!&0\!&0\!&0&1\\
0&0&0&0&0
\end{array}\right]\left[\begin{array}{c}\delta \hat{x}_1(t)\\\delta \hat{x}_2(t)\\\delta \hat{x}_3(t)\\ \delta \hat{b}(t)\\\hat{d}(t)\end{array}\right]\\&~~~+
\left[\begin{array}{ccccc}
-1.34&-1.34&-1.34&0&0\\
-0.74&-0.74&-0.74&0&0\\
-0.51&-0.51&-0.51&0&0\\
20&20&20&0&0\\
0&0&0&0&0
\end{array}\right]\left[\begin{array}{c}\delta \hat{x}_1(t-0.025)\\\delta \hat{x}_2(t-0.05)\\\delta \hat{x}_3(t-0.075)\\ \delta \hat{b}(t)\\ \hat{d}(t)\end{array}\right]\\&~~~+\left[\begin{array}{ccc}
-970&0&0\\
0&-959&0\\
0&0&-956\\
0&0&0\\
0&0&0
\end{array}\right]\left[\begin{array}{c}\delta p_1(t-0.)\\ \delta p_2(t-0.) \\ \delta p_3(t-0)\end{array}\right]\\
&~~~+L\left[\begin{array}{ccccc}0&0&0&1&0\end{array}\right]\left(\tilde{x}(t)-\hat{x}(t)\right)
\end{aligned}
\end{equation}}
where the observer gain $L$ ensures the convergence of $\hat{x}(t)$ to the real state $\tilde{x}$. Applying Theorem \ref{theorem2}, such a matrix gain can be found: $L=\left[0.28~0.46~0.45~1.76~0.54\right]^T$. Prior theoritical simulations with the non linear model (\ref{modelNL2}) under Matlab/Simulink show that the mechanism works well (see Figure \ref{matlab}). Then, we have performed a simulation of $400s$ on NS-2 where the $20$ ftp connections of each three TCP sources send data to their respective receivers. An additional non responsive traffic generated by $3$ UDP (user datagram protocol) traffic (at $1Mbps$ each one) is injected into the bottleneck as illustrated in Figure \ref{topologie3}. This latter simulates a CBR anomaly and is introduced at intervals: $150-170$s, $250-270$s and $300-320$s.\\
~\indent Estimation of the state and instantaneous measures are compared (the queue length and sending rates) as well as the anomaly detection \textquotedblleft sensor\textquotedblright  ~is illustrated in Figure \ref{observer_K}. Results show that reconstructing the state of model (\ref{forme_can}), the time-delay observer (\ref{observer}) is able to provide an estimation of TCP flow rates only based on the queue length measurement. Furthermore, the augmented model (\ref{augmented-model}) allows the observer to detect also non-modeled piecewise constant traffic. Hence, as it can be seen in Figure \ref{observer_K}, although the anomaly does not affect the queue (this attack is invisible from the buffer measurement), the mechanism can clearly detect the three UDP anomalies.


\begin{remark}
Regarding to NS-2 simulations, Figure \ref{observer_K}, the estimated state follows the linearized model (\ref{modelNL2}) which considers the network mean variables whereas the original state measured in NS gives instantaneous values. That is why such large oscillations around estimated signals are obtained for the measurements.
\end{remark}

Table \ref{tab} shows that  the observer state matches average flow rates.
\begin{table}
\label{tab}
  \centering
  \caption{Average of measure/estimated of flow rates}\label{stats3}
\begin{tabular}{|c|c|c|}
  \hline
  & Simulation& Emulation \\
  \begin{tabular}{c}
   \!\!\!\!\!\!\!\! \\\!\!\!\! $x_1(t)$ (pkt/s)\!\!\!\! \\\!\!\!\! $x_2(t)$ (pkt/s)\!\!\!\!  \\ \!\!\!\! $x_3(t)$ (pkt/s) \!\!\!\!
    \end{tabular}
    & \!\!\!\! \begin{tabular}{c|c}measured&estimated\\ 51&43 \\49&43  \\53&45 \end{tabular}\!\!\!\! & \!\!\!\! \begin{tabular}{c|c}measured&estimated\\ 92&99 \\93&100 \\113&110 \end{tabular}\!\!\!\!  \\
  \hline
\end{tabular}
\end{table}

\subsection{Emulation}

\begin{figure}
\vspace*{-1cm}
       \centerline{\includegraphics[height=7cm, width=8cm]{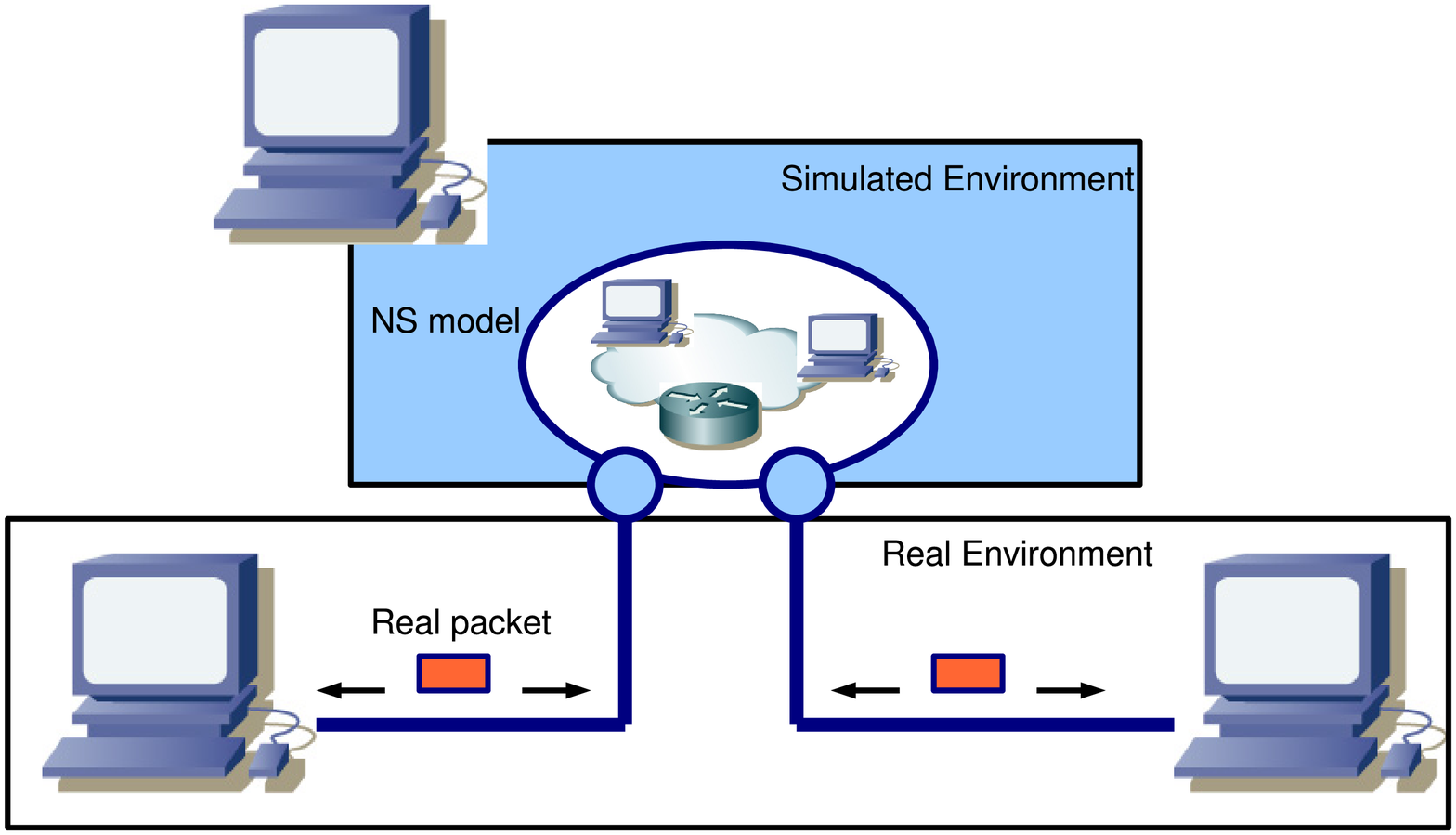}}
       \caption{Emulation with real-time NS}
       \label{NSemulation}
\end{figure}

Going further than simulations, another example is proposed considering now emulation experiment. Emulation refers to experiments that introduce the simulator into a live network. Indeed, the NS environment provides special objects that allow the simulator to interact (catch and inject) real traffics using a real-time scheduler (see \cite{Fal99} and Figure \ref{NSemulation}).\\
~\indent Regarding our study, the NS environment will be embedded in the computer that plays the role of the routers and other computers will generate and receive the traffic. Hence, the bottleneck and the observer are emulated while a real TCP traffic is handled and monitored.

\begin{figure}
       \centerline{\includegraphics[height=6cm, width=8cm]{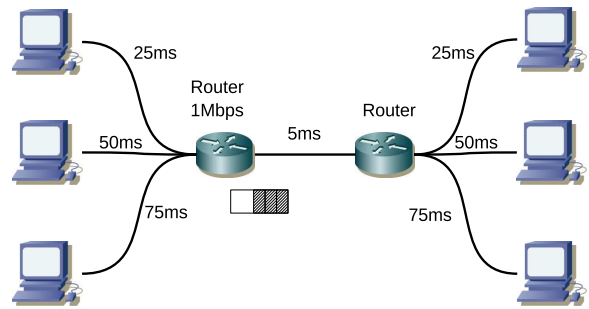}}
\vspace*{-1.5cm}
       \caption{Second example of a bottleneck link}
       \label{exemple2}
\end{figure}

\begin{figure}
       \centerline{\includegraphics[height=5cm, width=8cm]{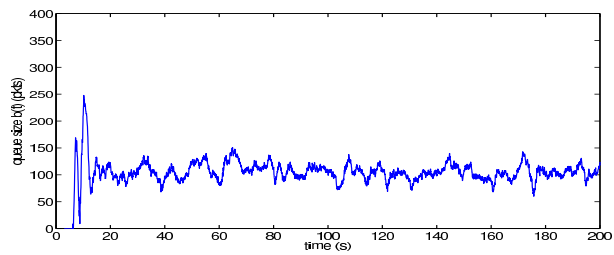}}
       \caption{Observer over gain-K: regulation of the queue length}
       \label{queueemule}
\end{figure}

\begin{figure}
       \centerline{\includegraphics[height=12cm, width=9.5cm]{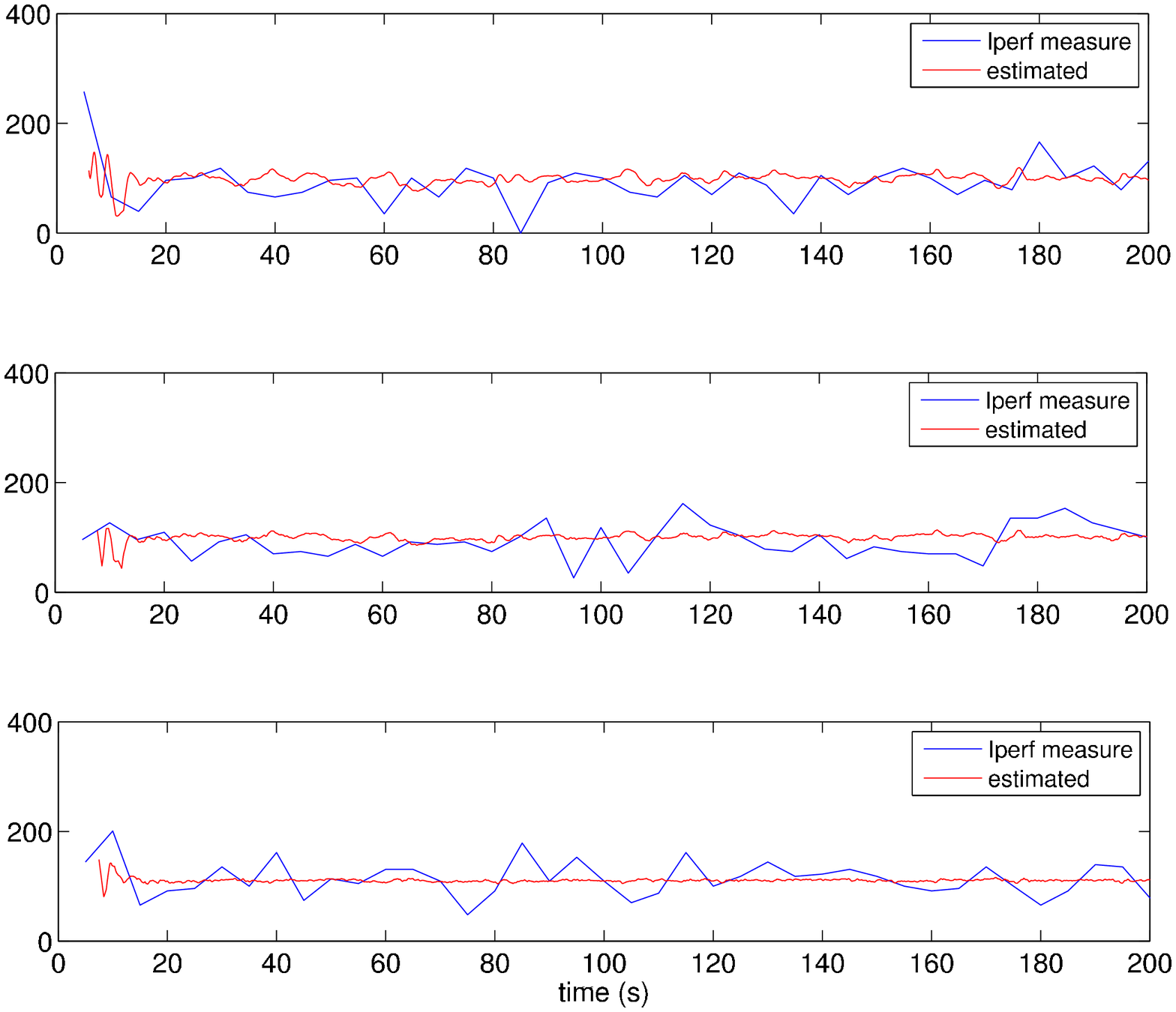}}
\vspace*{-2cm}
       \caption{Observer over gain-K: original/estimated of rates}
       \label{emule}
\end{figure}

However, since the emulator requires a high computational cost, numerical values of the example must be scaled down (reducing the bandwidth). The considered example is illustrated in Figure \ref{exemple2}. Source traffics are generated with the network tool Iperf \cite{iperf}. Applying a congestion control mechanism, the queue size of the buffer is regulated (see Figure \ref{queueemule}) and a linear observer can thus be developed according to the appropriate equilibrium point. Results of the emulation are shown in Figure \ref{emule}.

~\indent As it has been noticed in Remark 1, the state of the observer corresponds to average rates (see Table \ref{tab}). Because of the network load (3 heavy data streams by source which cause congestion phenomenon), Iperf gives measures of the rate at a sampling period of $5s$. That is why, in Figure \ref{emule}, measures appear dispersed around the estimation.

~\indent A network emulator may be considered as an hybrid between a network simulator and a protocol implementation. Future works concern the real implementation of the AQM/observer into the Linux kernel to enable whole real experiments and real traffic monitoring in high speed networks.\\

\section{CONCLUSIONS AND FUTURE WORKS}
~\indent In this paper, robust control theory tools have been used to design an observer for traffic monitoring purpose. This latter is embedded in a router and provides TCP flows estimations which pass through it. However, since the proposed observer is linear, an AQM that regulates the traffic around an equilibrium point is required. Besides, an augmented model is developed and the associated observer allows the detection of a class of anomalies in order to prevent potential malicious traffic as DoS attacks.\\
~\indent Future works concerns modeling studies of other existing DoS or DDoS attacks to endow the observer (by model augmentation) of a larger versatile anomaly detection system. Another point is the development of a non linear observer able to reconstruct the state, thus the traffic, without the AQM requirement.


\bibliographystyle{plain}
\bibliography{mabiblio}

\end{document}